\documentstyle[preprint,pra,aps]{revtex}

\begin{document}

\draft
%\twocolumn
\title{Information-theoretic aspects of quantum inseparability of mixed
states}

\author{Ryszard Horodecki\cite{poczta}}

\address{Institute of Theoretical Physics and Astrophysics\\
University of Gda\'nsk, 80--952 Gda\'nsk, Poland}

\author{ Micha\l{} Horodecki}

\address{Departament of Mathematics and Physics\\
 University of Gda\'nsk, 80--952 Gda\'nsk, Poland}

\maketitle

\begin{abstract}
Information-theoretic aspects of quantum inseparability of mixed states 
are investigated in terms of the $\alpha$-entropy inequalities and 
teleportation fidelity.  Inseparability of mixed states is defined and 
a complete characterization of the inseparable $2\times2$ systems with 
maximally disordered subsystems is presented within the 
Hilbert-Schmidt space formalism. A connection between  teleportation 
and negative conditional $\alpha$-entropy is also emphasized.
\end{abstract}

\section{Introduction}

Quantum inseparability is one of the most striking features of quantum
formalism. It can be expressed as follows:
{\it
If two	systems interacted in the past it is,  in general, not possible to
assign a single state vector to either of the to subsystems.
} \cite{blum,desp}. This is what is sometimes called the principle of
inseparability. Historically it was first recognized by Einstein, Podolsky and
Rosen
(EPR) \cite{epr} and by Schr\"odinger \cite{schnat}. In their famous paper EPR
suggested a description of the world (called ``local realism'') which assigns
an independent and objective reality to the physical properties of the well
separated substystems of a compound system.
Then EPR used the criterion of local realism to conclude that quantum mechanics
is incomplete.

EPR criticism was the source of many discussions concerning fundamental
differences between quantum and classical description of nature.
The most significant progress toward the resolution of the EPR problem was
made by Bell \cite{bell} who proved
that the local realism implies constraints
on the predictions of spin correlations in the form of inequalities
(called Bell's inequalities) which can be violated by some quantum mechanical
predictions.
The latter feature of quantum mechanics called usually ``nonlocality''
\cite{nonlocality}
is one of the most apparent manifestations of quantum inseparability.

The Bell's inequalities involve correlations between the
outcomes of measurements performed on the well separated systems which have
interacted in the past. It emphasizes the correlation aspect of
inseparability. There is another aspect which cannot be directly related to
correlations but rather to amount of information carried by quantum states.
It was first considered by Schr\"odinger who wrote in the context of the EPR
problem:
``Thus one disposes provisionally (until the entanglement is resolved by actual
observation) of only a {\it common} description of the two in that space of
higher dimension. This is the reason that knowledge of the individual systems
can {\it decline} to the scantiest, even to zero, while that of the combined
system remains continually maximal. Best possible knowledge of a whole
does {\it not} include best possible knowledge of its parts -- and that is
what keeps coming back to haunt us'' \cite{schnat}. In this way Schr\"odinger
recognized a
profoundly nonclassical relation between the information which an entangled
state gives us about the whole system and the information which it gives us
about the subsystems. It involves an information-theoretic aspect of quantum
inseparability which
has attracted much attention recently \cite{caves,schum,barph,inf,red,reney}.
Braunstein and Caves first
considered information-theoretic Bell's inequalities, and have shown
that they can be violated in the region of the violation of the usual Bell's
inequalities \cite{caves}(see also Ref. \cite{schum} in this context).
There is another approach  which bases on the notion
of  the index of correlation  \cite{barph,inf} or, more generally, quantum
redundancies \cite{red}. In particular, it has been shown \cite{red} that for
all known states admitting the local hidden variable (LHV) model the
normalized index of correlation is bounded by $1\over2$ \cite{red}. More
general analysis in terms of the so-called $\alpha$-entropy inequalities
\cite{reney} shows that there is a connection between the  correlation aspect
and the information-theoretic one involving quantum entropies.

Recently Bennett at al. \cite{bennett} have discovered a new
aspect of quantum inseparability  --
teleportation.
It involves a separation of an input state into classical and quantum part
from which the state can be reconstructed with perfect fidelity ${\cal F}=1$.
The basic idea is to use a pair of particles in the singlet state shared by the
sender (Alice) and the receiver (Bob). Popescu \cite{pop1} noticed that the
pairs in a mixed state  could be still useful for (imperfect) teleportation.
There was
a question what value of the fidelity of transmission of an unknown state can
ensure us about nonclassical character of the state forming the quantum
channel. It has been shown \cite{pop1,massar} that  purely classical channel
can give at most ${\cal F}={2\over3}$.  Subsequently, Popescu showed that
there exist mixtures which are useful for teleportation although they admit
the local hidden variable (LHV)  model.
Then basic
questions concerning the possible relations between  teleportation, Bell's
inequalities and inseparability were addressed \cite{pop1}.
Quite recently the maximal fidelity for the standard teleportation scheme
\cite{stand} with the quantum channel formed by any mixed two spin-$1\over2$
state has been obtained \cite{telep}.

The main purpose of the present paper is to investigate a relation
between  inseparability of mixed states and their
nonclassical information-theoretic features. In particular
we provide  a complete  characterization of the inseparable
$2\times2$ systems with maximally disordered subsystems.
This paper is organized in the following way. In sec. \ref{prin} we discuss the
inseparability principle for mixed states. In sec. \ref{entr} we present
the quantum $\alpha$-entropy ($\alpha$-E) inequalities and discuss them in the
context of inseparability. In particular, we point out that
separable states satisfy $1$ and $2$-entropy inequalities.
In sec. \ref{hs}  we consider
in detail $2\times 2$ systems using the Hilbert-Schmidt space formalism.
In particular, we provide the necessary conditions for separability of the
mixed two spin-$1\over2$ states.
In sec. \ref{maxy}
we provide a complete characterization of the mixed two spin-$1\over2$ states
with maximal entropies of subsystems, and we single out the separable states
belonging to the above class. This allows us to obtain
information-theoretic characterizations of the latter in terms of
the $\alpha$-entropy inequalities and teleportation presented in sec.
\ref{char}. The characterization is also obtained in terms of purification of
noisy teleportation channels \cite{bpop}. Finally we discuss the idea of
purification in the context of $\alpha$-E inequalities.

\section{Inseparability principle} \label{prin}
To make our consideration more clear it is necessary to extend the notion of
inseparability.
Note that the above principle of inseparability  when
applied to mixed states becomes inadequate. Indeed,
there are nonproduct mixtures that can be written as mixtures of  pure
product states thus  separable according to the principle. Then for clarity
it is convenient to introduce the following natural generalization
of the latter:
{\it If two systems interacted in the past it
is possible to find the whole system in the state that cannot be written as a
mixture of product states}.
It involves the existence of inseparable mixed states which may be viewed
as a counterpart of  pure entangled states.
They correspond to the Werner's EPR correlated
states i.e. the ones which cannot be written as mixtures of direct products,
while the separable states (mixtures of product states) correspond to the
classically  correlated ones \cite{werner}. It is natural to interprete the
principle as follows.
As one knows, a mixed state  can in general come from reduction of
some pure state or from a source producing randomly pure states. If a
mixed state is separable, then it produces the statistics equivalent to the
one generated by an ensemble of product states. In the latter case, the
nonfactorability is  due to the lack of knowledge of the observer only. However
if a mixed state is inseparable, then there is certainly no way to ascribe
to the subsystems, even in principle, their state vectors.

Now, we are interested in information-theoretic aspects of
inseparability
as well as in the range of their manifestations. The question is also,
to what degree this range can cover the whole set of  inseparable states.
However it is very difficult to check whether some given
state can be written as a mixture of product states or not. Then it follows
that more ``operational'' characterization of inseparable
(separable) states is more than desirable.

It should be emphasized here that, although
the above inseparability principle
says  about  the existence of the dynamics that can convert product state into
inseparable one, we are interested in the
final effect of the action of the dynamics. In other words, we assume that the
system is found in an inseparable state, and the task is to investigate the
effects that manifest its inseparability.

\section{$\alpha$-entropy inequalities}\label{entr}
In this section we will outline the concept of the $\alpha$-entropy
 inequalities in the context of inseparability \cite{reney}.
Let us consider the
quantum counterpart of the R\'enyi $\alpha$-entropy \cite{thir,wehrl}
\begin{equation}
S_\alpha(\varrho)={1\over 1-\alpha} \ln Tr\varrho^\alpha,\quad\alpha>1.
\label{ogol}
\end{equation}
If $\alpha$ tends  to $1$ decreasingly, one obtains the von Neumann entropy
$S_1(\varrho)$ as a limitting case
\begin{equation}
S_1(\varrho)=-Tr\varrho\ln\varrho.
\end{equation}
%The $\alpha$-entropies could be used as a measure of the information content
%of the
One can replace the standard information measure which is von Neumann
entropy by the whole  family of $\alpha$-entropies \cite{graud}.
Then given the compound system, one can consider the relationships between the
entropy of the whole system and the entropies of the subsystems. For this
purpose, consider the following inequalities
\begin{equation}
S_\alpha(\varrho)\geq\max_{i=1,2}S_\alpha(\varrho_i),
\label{ren}
\end{equation}
where $\alpha\geq1$, $S_\alpha(\varrho)$ denotes the entropy of the system
and $S_\alpha(\varrho_i),\ i=1,2$ are the entropies of the subsystems.
The above inequalities can be interpreted as the constraints imposed on the
system by positivity of the conditional $\alpha$-entropies if the latter
are defined by
\begin{equation}
S_\alpha(1|2)=S_\alpha(\varrho)-S_\alpha(\varrho_2),\quad
S_\alpha(2|1)=S_\alpha(\varrho)-S_\alpha(\varrho_1).
\end{equation}
Now one  can expect that violation of the $\alpha$-E inequalities is a
manifestation of some nonclassical features of a compound system
resulting from its inseparability. Indeed, one can easily see that
for discrete classical systems  \cite{disc} the corresponding inequalities
are always satisfied i.e. the classical conditional $\alpha$-entropies
are positive.

Note that the classical systems are always separable: joint distributions can
always be written as convex combination of product distributions
\cite{werner}. Then it is natural to ask about the connection between the
violation of the $\alpha$-E inequalities and inseparability of quantum
states. In fact one can prove \cite{reney}

\noindent {\bf Theorem 1 }
\rm For any separable state $\varrho$ on
the finite dimensional Hilbert space the inequality (\ref{ren})
is satisfied for $\alpha=1,2$.

\noindent
The above theorem provides   necessary conditions for separability.
In particular, it turns out that the $2$-E inequality is essentially
stronger than the Bell-CHSH inequality \cite{reney}. Then it constitutes
a nontrivial and extremely simple computationally necessary condition for 
separability.
This is a useful result as  there is still no operational
criterion of inseparability, in general. As we will see further,
for a class of two spin-$1\over2$ states, it is possible to provide the
criterion in terms of the $\alpha$-E inequalities.

\section{Positivity and separability conditions in the Hilbert-Schmidt space}
\label{hs}
In this section and further we will restrict our consideration to the
$2\times2$ systems. Consequently, consider the Hilbert space
${\cal H}=C^2\otimes C^2$.
All Hermitian operators acting on ${\cal H}$ constitute a Hil\-bert-Schmidt
(H-S) space
${\cal H}_{\text{H-S}}$  with a scalar product
$\langle A,B\rangle=Tr(A^\dagger B)$.
An arbitrary state of the system can be represented in
${\cal H}_{\text{H-S}}$ as follows
\begin{equation}
\varrho={1\over4}(I\otimes I+\bbox {r\cdot\sigma}\otimes I+I\otimes\bbox
{s\cdot \sigma}+
\sum_{m,n=1}^3t_{nm}\sigma_{n}\otimes\sigma_{m}).
\label{postac}
\end{equation}
Here
$I$ stands for identity operator,
$ {\bbox r}$, ${\bbox s}$ belong to $R^3$,
$\{\sigma_n\}^3_{n=1}$ are the standard Pauli
matrices, $\bbox{ r\cdot\sigma}=\sum_{i=1}^3 r_{i}\sigma_{i} $.
The coefficients $t_{mn}={\rm Tr}(\rho\sigma_n \otimes \sigma_m)$
form a real matrix denoted by $T$.
Note that $\bbox r$ and $\bbox s $ are local parameters as they determine
the reductions of the state~$\varrho$
\begin{eqnarray}
\varrho_1\equiv{\rm {\rm Tr}}_{{\cal H}_2}\varrho={1\over2}(I+\bbox{
r\cdot\sigma}),\nonumber\\
\varrho_2\equiv{\rm {\rm Tr}}_{{\cal H}_1}\varrho={1\over2}(I+
\bbox{s\cdot\sigma}).
\label{redukcje}
\end{eqnarray}
while the $T$ matrix is responsible for correlations
\begin{equation}
E(\bbox{a},\bbox{b})
\equiv \text{Tr}(\varrho\bbox{a\cdot\sigma}\otimes\bbox{b\cdot\sigma})=(\bbox
a,T\bbox b).
\end{equation}
%Note that the form (\ref{postac}) does not include nonnegativity condition.
Now we will reduce the number of the parameters that are essential for the
problem we discuss in this paper. Note that inseparability is invariant under
product unitary transformations i.e. if a state $\varrho$ is inseparable
(separable) then any state of the form $U_1\otimes U_2\varrho U_1^\dagger
\otimes U_2^\dagger$ also have this property. Then, without loss of generality
we can restrict our considerations to some representative class of the states
described by less number of parameters. The class should be representative in
the sense that any state $\varrho$ should be of the form
$\varrho=U_1\otimes U_2\tilde\varrho U_1^\dagger\otimes U_2^\dagger$
where $\tilde\varrho$ belongs to the class.
Consequently, denote by ${\cal D}$  the set of all states with diagonal $T$.
This set is a convex subset of the set of all states. To show
that it is representative, we will use the fact, that for any unitary
transformation $U$ there is a unique rotation $O$ such that \cite{thir2}
\begin{equation}
U\bbox{ \hat n\cdot\sigma}U^\dagger
=(O\bbox{\hat n})\bbox{\cdot\sigma}.
\label{pawel}
\end{equation}

Then if a state is subjected to the $U_1\otimes U_2$ transformation,
 the parameters $\bbox r, \bbox s$  and  $T$ transform
themselves as follows
\begin{eqnarray}
&&\bbox r'=O_1\bbox r,\nonumber\\
&&\bbox s'=O_2\bbox s,\nonumber\\
&&T'=O_1 T O_2^\dagger.\nonumber
\end{eqnarray}
where $O_i$'s correspond to $U_i$'s via formula (\ref{pawel}).
Thus, given an arbitrary state, we can always choose such $U_1, U_2$ that the 
corresponding rotations will diagonalize its matrix $T$ \cite{perf},
 and the transformed
state will belong to $\cal D$.

Now we are in position to present the conditions imposed on the parameters
contained in the $T$ matrix by positivity of $\varrho$ and by its
separability \cite{telep}.  
As we consider the states with diagonal $T$ we can identify the
latter with the vector $\bbox t\in R^3$ given by $\bbox t=
(t_{11},t_{22},t_{33})$. Henceforth we will identify diagonal matrices with
corresponding vectors. By the notation $T\in {\cal A}$, where $T$ is a
matrix and $\cal A$ is a subset of $R^3$, we mean that $T$ is diagonal and
the corresponding vector belongs to $\cal A$.
We have %\cite{telep}
%\begin{prop}

\noindent {\bf Prop. 1 }{\it
For any $\varrho\in{\cal D}$ the $T$ matrix  given by (\ref{postac}) belongs 
to the tetrahedron $\cal T$ with vertices $\bbox t_0=(-1,-1,-1)$,  
$\bbox t_1=(-1,1,1)$, $\bbox t_2=(1,-1,1)$, $\bbox t_3=(1,1,-1)$.}
%\label{positivity}

\noindent
{\it Proof\/}.- $\varrho$ is positive iff the following  inequalities are
satisfied
\begin{equation}
\text{Tr}(\varrho P)\geq 0,
\end{equation}
for any projector $P$. Consider four projectors given by Bell basis
\begin{eqnarray}
&&\psi_{1\atop(2)}={1\over\sqrt2}(e_1\otimes e_1\pm e_2\otimes e_2)\\
&&\psi_{3\atop(4)}={1\over\sqrt2}(e_1\otimes e_2\pm e_2\otimes e_1).
\label{baza}
\end{eqnarray}
where $\{e_i\}$ is the standard basis in $C^2$. The parameters of the above
projectors in the Hilbert-Schmidt space are
\begin{eqnarray}
&&\bbox r_i=0, \ \bbox s_i=0,\quad i=0,1,2,3\nonumber\\
&&T_0=\text{diag}(-1,-1,-1)\nonumber\\
&&T_1=\text{diag}(-1, 1, 1)\nonumber\\
&&T_2=\text{diag}( 1,-1, 1)\nonumber\\
&&T_3=\text{diag}( 1, 1,-1).
\label{parbazy}
\end{eqnarray}
Now as for two states $\varrho$ and $\varrho'$ given by
$(\bbox r,\bbox s, T)$ and $(\bbox r',\bbox s', T')$ respectively, one has
\begin{equation}
\text{Tr}\varrho \varrho'={1\over4}(1+(\bbox r,\bbox r')+(\bbox s,\bbox s')+
\text{Tr}(TT'^\dagger))
\end{equation}
Then one obtains that the inequalities $\text{Tr}\varrho P_i\geq 0$,
$i=0,1,2,3$
are equivalent to the following ones
\begin{eqnarray}
1-t_{11}-t_{22}-t_{33}\geq0\nonumber\\
1-t_{11}+t_{22}+t_{33}\geq0\nonumber\\
1+t_{11}-t_{22}+t_{33}\geq0\nonumber\\
1+t_{11}+t_{22}-t_{33}\geq0\nonumber.
\end{eqnarray}
Clearly, the above conditions mean that $T$ belongs to the tetrahedron
$\cal T$.

Now one can  establish conditions implied by separability of a given
state $\varrho$. Consequently, we have

\noindent {\bf Prop. 2 }{\it
%\begin{prop}
For any separable state $\varrho\in{\cal D}$ the $T$ matrix given by
(\ref{postac}) belongs to the octahedron $\cal L$ with vertices
$\bbox o^{\pm}_1=(0,0,\pm1)$,
$\bbox o^{\pm}_2=(0,\pm1,0)$,
$\bbox o^{\pm}_3=(0,0,\pm1)$.}
%\label{separability}
%\end{prop}

\noindent
{\it Proof\/}.- Consider the operator $V$ given by
$V\phi\otimes\tilde\phi=\tilde\phi\otimes\phi$. Note that \cite{werner}
one has
\begin{equation}
\text{Tr}VA\otimes B=\text{Tr}AB
\end{equation}
Then it follows that for any separable state $\varrho$ we have
\begin{equation}
\text{Tr}V\varrho\geq0.
\label{wer}
\end{equation}
Consider now four operators $V_i$, $i=0,1,2,3$ given by
\begin{equation}
V_i=\sigma_i V\sigma_i
\end{equation}
where $\sigma_0\equiv I$.
As the set of separable states is $U_1\otimes U_2$ invariant, it follows from
(\ref{wer}) that for separable state $\varrho$ we have
\begin{equation}
\text{Tr}V_i\varrho\geq0, \quad i=0,1,2,3.
\label{wer2}
\end{equation}
The operators $V_i$ can be written in terms of the H-S space
\begin{equation}
V_i={1\over4}(2 I\otimes I - \sum_{j=1}^3 t^{i}_{jj}\sigma_j\otimes\sigma_j)
\end{equation}
where $T^{i}\equiv\text{diag}(t^{i}_{11},t^{i}_{22},t^{i}_{33})=-2T_i$.
Then the inequalities (\ref{wer2}) imply  $T\in-\cal T$
($\bbox{t}\in-{\cal T}$ iff $-\bbox{t}\in{\cal T}$). Combining this with
Prop. 1 we obtain that  for
separable states with diagonal $T$, the latter  belongs to the ocathedron
$\cal L={\cal T}\cap-{\cal T}$.

\section{$2\times2$ systems with maximally disordered subsystems}
\label{maxy}
In this section we shall deal with the states the reduced density matrices of
which are maximally disordered i.e. are normalized identities.
The only pure states with this property are maximally entangled ones
i.e.  $U_1\otimes U_2$ transformations of the singlet state.
The latter appears to be the most nonclassical of all pure states.
Many of  mixtures belonging to the class of the
states with maximally disordered subsystems should exhibit nonclassical
properties as, if the entropies of the subsystems are maximal, we expect that
the inequalities (\ref{ren}) should often be violated.

The states with maximally disordered subsystems are completely
characterized by $T$ matrix (we will call them further the $T$-states).
Again, we can restrict our considerations to the states with diagonal $T$.
We prove here

\noindent {\bf Prop. 3 }{\it
Any operator of the form (\ref{postac}) with $\bbox r,\bbox s=0$ and diagonal
$T$ is a state iff $T$ belongs to the tetrahedron $\cal T$.}

\noindent
{\it Proof\/}.- If the operator is a state, then from Prop. 1
it follows that $T$ belongs to the tetrahedron. Now, let $T$ belong to the
tetrahedron. Then it can be written as a convex  combination of its vertices
treated as matrices. Thus the operator given by $T$ appears to be a convex
combination of the projectors given by (\ref{baza}).

Note that the necessary condition of positivity of the operators in the H-S
space given by Prop. 1 is now also sufficient.
The above proposition gives us complete characterization of the states with
maximal entropies of the subsystems. Any such state is of the form:
\begin{equation}
\varrho=U_1\otimes U_2(\sum_{i=0}^3p_iP_i)U_1^\dagger\otimes U_2^\dagger
\label{max}
\end{equation}
where $U_1, U_2$ are unitary transformations, $P_i$'s are given by
(\ref{baza}) and $\sum_ip_i=1$.
Thus one can see that the above class appears to be
a very poor one: up to the $U_1\otimes U_2$ isomorphism, the states have only
one partition of unity (in the nondegenerate case).

Now it is important to know  which of the considered states are inseparable.
It turns out that, again, the necessary condition for separability of the
states in the H-S space given by Prop. 2 appears to be
sufficient in the case of the considered states.
Namely we have

\noindent {\bf Prop. 4 }{\it
Any two spin-$1\over2$ state with maximally disordered subsystems and diagonal
$T$ is separable iff $T$ belongs to the octahedron $\cal L$.}
%\label{sep}

\noindent
{\it Proof\/}.- To prove sufficiency, note that the vertices $o^{\pm}_k$,
$k=1,2,3$, of the octahedron
represent some separable states. Indeed, one can easily check that they
represent the states
\begin{equation}
\varrho^{{+\atop(-)}}_k={1\over 2}(P_k^{{+\atop(+)}}\otimes P_k^{{+\atop(-)}}+
P_k^{{-\atop(-)}}\otimes P_k^{{-\atop(+)}}),\quad k=1,2,3.
\label{roka}
\end{equation}
where $P_k^\pm$ correspond to the eigenvectors of $\sigma_k$ with
eigenvalues $\pm1$.
Now if $T$ belongs to the octahedron, it can be written as a convex
combination of its vertices. Thus the corresponding state can be written
as a mixture of the states $\varrho^\pm_k$, hence is a separable state.

It is remarkable  that the above result can easily be expressed in terms
of spectra of the
considered states. Indeed, the octahedron represents the $T$-states
with diagonal $T$ that have all
the eigenvalues less than or equal to $1\over2$. As the spectrum is invariant
under unitary transformations (then $U_1\otimes U_2$ invariant) we obtain

\noindent {\it Prop. 5} {\it
Any state $\varrho$ with maximal entropy of the subsystems is separable iff
$\sigma\subset[0,{1\over2}]$ where $\sigma$ is the spectrum of $\varrho$.}

\section{Inseparable $2\times2$  systems with maximally disordered
subsystems: information-theoretic characterizations}
\label{char}
\subsection{$\alpha$-entropy characterization}
There is an interesting  relation between inseparability of the $T$-states and
their nonclassical information-theoretic features.
It turns out that the condition imposed on the spectrum of the
states can be expressed as the amount of the classical conditions for
the $\alpha$-entropies. Namely we have

\noindent {\bf Prop. 6 }{\it
A state with maximal entropies of subsystems is separable iff it satisfies
the $\alpha$-E inequalities for all $\alpha\geq1$.}

\noindent
{\it Proof\/}.- A simple proof is based on geometrical arguments.
For the $T$-states the $\alpha$-E inequalities read
\begin{eqnarray}
&&\sum_ip_i^\alpha\leq2^{1-\alpha}, \quad\alpha>1\nonumber\\
&&-\sum_ip_i\ln p_i\geq\ln2, \quad\alpha=1
\end{eqnarray}
where $\{p_i\}$ is  the spectrum. Thus the set of distributions
$\{p_i\}$ satisfying the inequalities is convex, hence the subset $\cal E$
of the $T$-states with diagonal $T$ satisfying them is
also convex.
Now, as the states $\varrho^{\pm}_k$ given by (\ref{roka}) satisfy the
$\alpha$-E inequalities, then all the states from the octahedron must also
satisfy them. To see that there are no states beyond the octahedron with
this property,
it suffices to consider the states represented by the line connecting one of
the vertices of $\cal T$ (e.g. the one representing the singlet state
$\psi_0$) with
the origin of the frame \cite{werpop} (see Fig. \ref{rys}). It is
straightforward to check that a
state belonging to this line satisfies the $\alpha$-E inequalities iff it
is separable \cite{reney} (i.e. iff it belongs to the octahedron).

Now from $U_1\otimes U_2$ invariance of the set $\cal E$ it follows that
it is invariant under the group of proper rotations of a regular tetrahedron.
Then from the convexity of the considered set it follows that it must be the
octahedron i.e. we have ${\cal E}={\cal L}$.

Thus we see that the inseparability of the $2\times2$ states with maximal
entropies of the reductions manifests itself by violation of the
$\alpha$-entropy
inequalities for some $\alpha$ i.e. by negativity of some conditional
$\alpha$-entropy \cite{large}.
Recently the negative conditional von Neumann entropy was considered in the
context of the teleportation and superdense coding \cite{eprint}.
As we will see further, within the considered class of the states,
the negativity of any conditional $\alpha$-entropy makes the state useful for
teleportation.

\subsection{Teleportation characterization}
Some inseparable states have a very striking feature. Namely they can be used
for transmission of quantum information with
better fidelity than by means of classical bits themselves. For example,
two-particle
system in pure singlet state shared by a sender and a receiver allows to
transmit faithfully an unknown two spin $1\over2$ state, with additional use
of two classical bits \cite{bennett}. This is what  is called quantum
teleportation. In absence of the quantum channel, the only thing
the sender can do is to measure the unknown state
and then to inform the receiver about the outcome of the measurement by means
of classical bits.
Now, if in presence of the quantum channel the receiver can reconstruct
the state better than it is possible by using the best possible strategy
basing only on classical bits, then  we say that the state forming the quantum
channel is useful for teleportation \cite{pop1}.

Then there is a basic question: are all the inseparable
$T$-states useful for teleportation? As we will see below, the answer
is ``yes''. As a measure of efficiency of  teleportation we will use fidelity
\begin{equation}
{\cal F}=\int_S\text dM(\phi)\sum_kp_k\text{Tr}(\varrho_k P_\phi).
\label{fidelity}
\end{equation}
Here $P_\phi$ is the input state, $\varrho_k$ is the output state, provided the
outcome $k$ was obtained by Alice. The quantity $\text{Tr}(\varrho_k P_\phi)$
which is a measure of how the resulting state is similar to the input one, is
averaged over the probabilities of the outcomes, and then over all possible
input states ($M$ denotes uniform distribution on the Bloch sphere $S$).
It has been shown, that the purely classical channel can give at most
${\cal F}={2\over3}$ \cite{pop1,massar}.
Recently it has been proved \cite{telep} that within the standard teleportation
scheme \cite{stand} the inequality ${\cal F}\leq{2\over3}$ is equivalent to
the following one
\begin{equation}
N(\varrho)\leq1
\end{equation}
where $N(\varrho)\equiv\text{Tr}\sqrt{T^\dagger T}$.
Moreover, if a state is useful for the standard teleportation, then the maximal
fidelity amounts to
\begin{equation}
{\cal F}_{\max}={1\over2}(1+{1\over3}N(\varrho)).
\end{equation}
Now we observe that within the set $\cal D$ the inequality $N(\varrho)\leq 1$
holds iff $T$  belongs to the octahedron. Thus for the $T$-states with
diagonal $T$ this is equivalent to the separability condition
given by Prop. 4.  Obviously, if a state is separable then not only
the standard teleportation procedure but any possible one can not produce
fidelity greater than $2\over3$.
Then, under the $U_1\otimes U_2$ invariance of $N(\varrho)$ we
obtain

\noindent {\bf Prop. 7 }{\it
A two spin-$1\over2$ state with maximal entropies of the subsystems is useful
for teleportation iff it is inseparable.}

\noindent
Thus, again, inseparability of the $T$-states manifests itself inherently by
better fidelity of  teleportation than the maximal one produced by
purely classical channel. It is remarkable that, within the considered
class of states, the ability of forming efficient teleportation channel is due
to the negative conditional entropy for all large $\alpha$. This generalizes
earlier result concerning  Werner spin-$1\over2$ states
(see Fig. 1) \cite{reney}.

So far we have considered  teleportation directly via mixed states.
Recently, Bennett at al. \cite{bpop} presented the idea of purification of
noisy channels. The authors show how to obtain asymptotically faithful
teleportation  via mixed states using local operations and classical
communication in order to purify them. The state can be purified by
BBPSSW procedure  if
$\text{Tr}\varrho P_0>{1\over 2}$ where $P_0=|\psi_0\rangle\langle\psi_0|$ is
the singlet state. Of course,
one can immediately see that it can be also purified if $\text{Tr}\varrho
P>{1\over2}$, where $P$ is a projector corresponding to any maximally
ntangled pure state.
Thus by Prop. 5 we obtain that given a mixed $2\times2$ state with maximal
entropies of the subsystems, one can distill a nonzero entanglement by
using the BBPSSW procedure iff the state is inseparable.

%{\it Remark.-} Note that beyond the class of states considered in this section
%there are states which are inseparable, but still are not useful for
%standard teleportation. As an example consider the following class of states
%\cite{fazy,telep}
%\begin{equation}
%\varrho=p_1|\psi_1\rangle\langle\psi_1|+p_2|\psi_2\rangle\langle\psi_2|
%\label{stany}
%\end{equation}
%where
%\begin{eqnarray}
%|\psi_1\rangle =ae_1\otimes e_1+be_2\otimes e_2\\
%|\psi_2\rangle =ae_1\otimes e_2+be_2\otimes e_1
%\label{czyste}
%\end{eqnarray}
%with $a,b>0$, $\{e_i\}$ being standard basis in $C^2$,
%\ $0<(p_1-p_2)^2\leq(a^2-b^2)^2$.
%All the states (\ref{stany}) violate
%$\alpha$-E inequalities (which for two-component mixtures are all equivalent).
%Then they are all inseparable. However, we can choose the free parameters in
%such a way that many of the above states will have $N(\varrho)\leq1$. Then
%they also cannot be purified by direct application of the BBPSSW
%procedure. However, it turns out that there is a way to purify them.
%Indeed, one should combine the BBPSSW procedure with the Gisin's filtering
%method which has been use for revealing ``hidden nonlocality''
%\cite{gis,fazy}.
%Namely, without loss of generality we can assume $a>b$. Then a filter acting
%on the first particle and damping $e_1$ (not affecting $e_2$) can turn the
%initial states (\ref{stany}) to the ones of the same form but with
%$a=b={1\over \sqrt{2}}$. Such mixtures are $T$-states,
%and by Prop.\ref{sep} are inseparable. Then we can apply the BBPSSW procedure
%to obtain nonzero distillable entanglement.

\section{Conclusion}
We have investigated information-theoretic aspects of inseparability of
mixed states in terms of  
the $\alpha$-entropy
inequalities and teleportation. We have discussed some
general properties of the $\alpha$-E inequalities. Subsequently,
using the Hilbert-Schmidt space formalism we have provided the separability
conditions for $2\times 2$ systems. Then the set of the $T$-states  (the two
spin-$1\over2$ states with maximal entropies of the subsystems) has been
considered in detail.

It appears that, up to the $U_1\otimes U_2$ isomorphism, the set of
the $T$-states can be identified with some tetrahedron $\cal T$ in $R^3$,
whereas the separable $T$-states can be identified with an octahedron
contained in $\cal T$. The above, very illustrative  geometrical 
representation of
the both sets allowed to obtain information-theoretic interpretation of
inseparability of the considered states. Namely, it appears that the states
lying beyond the octahedron violate  $\alpha$-entropy
inequalities for all large $\alpha$.
The resulting negative conditional $\alpha$-entropy has its reflection in the
fact that all the inseparable $T$-states are useful for
teleportation. In addition, they have nonzero distillable entanglement, i.e.
one can use them for asymptotically faithful teleportation by using the
BBPSSW procedure.

Finally we believe that the results of the present paper can help in deeper
understanding of the connections between the inseparability and
the quantum information theory. Moreover, they may be also useful in the 
problem of the classification of mixed states under the nonlocality criterion
\cite{pop1,pop2,gis}.

\begin{figure}
\caption{Geometrical representation of the states with diagonal $T$ and
maximally disordered reduced density matrices: the bold-line-contoured
octahedron represents separable states, the dashed
line denotes the set of the Werner states with the singlet state $A$ and
normalized identity $E$.
Here $A=(-1,-1,-1)$, $B=(1,1,-1)$,
$C=(1,-1,1)$, $D=(-1,1,1)$. \label{rys}}
\end{figure}


\begin{references}
\bibitem{nonlocality}
The term ``nonlocality'' is slightly misleading, as in fact one deals
with violation of the conjunction of the two conditions: locality and
objectivism. However, with respect to common tradition, we will  use further
the term ``locality'' (``nonlocality'') related to satisfying
(violating) the assumptions of local realism respectively.
\bibitem{stand}
By standard teleportation scheme we mean here that Alice uses
Bell basis in her measurement while Bob is allowed to apply
any unitary transformations.
\bibitem{disc}
By the classical discrete states we mean here joint probability distributions
defined on product probability space $\Omega_1\times\Omega_2$ where
$\Omega_1=\{1,\ldots,n\}$, $\Omega_2=\{1,\ldots,m\}$
\bibitem{werpop}
The considered line represents  mixtures of singlet state and  the identity.
The mixtures belong to the family of the states introduced by Werner
\cite{werner} and they have been considered in the context of 
teleportation in Ref.  \cite{pop1}.
\bibitem{large}
In fact, we have violation of the $\alpha$-E inequalities not only for some
$\alpha$ but rather for all large $\alpha$. Indeed, for inseparable
Werner states the conditional $\alpha$-entropy is a decreasing function in
$\alpha$. On the other hand, for the states associated with the vertices of
octahedron, it is equal to zero for all $\alpha$. Moreover, within the set
of $T$-states with diagonal $T$ the function is convex.
Then it follows that the class
of the sets ${\cal E}_\alpha$ of the states satisfying $\alpha$-E inequality
is decreasing in $\alpha$. Hence if a $T$-state violates
$\alpha$-E inequality for some $\alpha$, then it violates all the $\alpha$-E
inequalities with greater $\alpha$.
\end{references}
\end{document}